# A Multiferroic Ceramic with Perovskite Structure: La$_{0.5}$Bi$_{0.5}$Mn$_{0.5}$Fe$_{0.5}$O$_{3.09}$

Asish K. Kundu[*], R. Ranjith, V. Pralong, N. Nguyen, B. Kundys, V. Caignaert, W. Prellier and B. Raveau

*Laboratoire CRISMAT, CNRS UMR 6508, ENSICAEN, 6 Bd Maréchal Juin, 14050 Caen Cedex 4, France*

ABO$_3$ perovskite multiferroic La$_{0.5}$Bi$_{0.5}$Mn$_{0.5}$Fe$_{0.5}$O$_{3.09}$ where the B-site cations is responsible for the magnetic properties and the A-site cation with lone pair electron is responsible for the ferroelectric properties was synthesized at normal conditions. This oxide exhibits a ferromagnetic transition around 240 K with a well defined hysteresis loop, and a significant reversible remnant polarization below 67K similar to ferroelectric behavior. The magnetic interaction is interpreted by the ferromagnetic Fe$^{3+}$-O-Mn$^{3+}$ and antiferromagnetic Fe$^{3+}$(Mn$^{3+}$)-O-Fe$^{3+}$(Mn$^{3+}$) interactions competed each other, whereas the ferroelectricity is predominantly due to the polar nature introduced by the 6s$^2$ lone pair of Bi$^{3+}$ cations.

*Corresponding author email: asish.k@gmail.com/asish.kundu@ensicaen.fr

Multiferroic oxides have attracted increasing attention in recent years due to their possible application towards storage materials and intriguing fundamental physics.[1] Among the naturally existing oxides, the presence of both ferromagnetism and ferroelectricity is a rare phenomenon. This phenomenon often occurs in perovskite-structure type-oxide having the general $ABO_3$ formula. The most well known examples of naturally existing are $BiFeO_3$[2] and $BiMnO_3$.[3] However, $BiFeO_3$ is a canted antiferromagnet, which gives rise to weak ferromagnetism and the later is metastable, requiring high pressure conditions for synthesis of bulk phases.[1-3] Recent studies on $Bi_2MnNiO_6$ throw light on synthesis of materials with one or more order parameters for realizing multiferroic properties or magnetoelectric effects.[4] Among most of the $Bi^{3+}$-based double perovskite system studied for multiferroic and magnetodielectric properties, they exhibit high sensitivity towards the B-site cationic ordering , but require high pressure conditions for synthesis.[4] In the process of exploration of a simple multiferroic perovskite the following facts are now well established; (i) Ferroelectricity (FE) and ferromagnetism are mutually exclusive due to the $d^0$ electronic structure of the B-element,[5] (ii) The occupation of different B-site cations with varying ionic radius provide an opportunity to realize a polar ground state[6] and (iii) The lattice distortion induced by cations with lone pair electrons such as $Pb^{2+}$ or $Bi^{3+}$ play a vital role on the FE properties as shown for $PbTiO_3$ in comparison with $BaTiO_3$.[1]

In our study, we consider the following points; (i) ferromagnetic (FM) perovskites, with strong FM interaction between the B-elements[2-3], (ii) an A-site cation containing a lone pair electron could induce ferroelectricity through lattice distortion[7] and (iii) able to synthesis at normal conditions through conventional and simple routes. Aforementioned points have motivated us to synthesize the compound $La_{0.5}Bi_{0.5}Mn_{0.5}Fe_{0.5}O_{3.09}$ and to exploit the possibility of multiferroic features. In this letter, the multiferroic properties of the single perovskite $La_{0.5}Bi_{0.5}Mn_{0.5}Fe_{0.5}O_{3.09}$ (LBMFO) are presented. We show that LBMFO compound satisfy the coexistence of one or more order parameters, but most importantly can be prepare under normal conditions of synthesis.

Single phase powders prepared by sol-gel method,[8] were ground and pressed in the form of parallelepiped bars and sintered in a platinum crucible at 1223 K for 24h in $O_2$ atmosphere. Refinements carried out by the Rietveld[9] method allowed the X-rays diffraction pattern to be indexed in an orthorhombic structure (*Pnma* space group, with *a*=5.534 Å, *b*=7.817 Å and *c*=5.556Å). The chemical analysis, made using the redox titration,[11] leads to a the stoechiometry composition of "$O_{3.09}$"

The $^{57}Fe$ transmission Mössbauer spectrum registered at room temperature, using a $^{57}Co/Rh$ source shows the presence of one iron Mössbauer site with isomer shift of 0.38(1) mm/s and a quadrupole splitting value $\Delta E$=0.55(1) mm/s. This confirms that iron is totally trivalent and in high-spin state, whereas manganese is consequently trivalent (small fraction may be in $Mn^{4+}$ due to excess oxygen)[9], in agreement with the previous results.[11]

Magnetization (M) measurement was carried out by a SQUID magnetometer (Quantum Design). The temperature-dependence of zero field cooled (ZFC) and field cooled (FC) magnetization curves for LBMFO under the applied fields of 100, 1000 and 5000 Oe showed a smooth paramagnetic to FM type transition ($T_C$) around 240 K (data not shown), similarly to $LaMn_{0.5}Fe_{0.5}O_3$.[8] At higher magnetic field (H>5000 Oe), the curves merge down to low temperature without a sign of magnetic saturation, even if the $T_C$ is prominent for all studied fields. The thermoremanent magnetization (TRM) studies (not shown here) also exhibited a feature similar to that of a spin or cluster glass type material.[11-12] The isothermal field-dependent magnetization curves, M(H), registered at different temperatures (Fig. 1a), show significant hysteresis loops below 240K, with unsaturated value of magnetization even at higher fields (up to 50 kOe). At 10K, the values of remanent magnetization and coercive field are 0.18 $\mu_B$/f.u. and 1.5 kOe, respectively. Although, in the present system weak FM transition is noticed, yet a well defined hysteresis loop is depicted below the transition temperature (at 235K, $H_C \sim$ 200 Oe). The magnetization behavior of LBMFO corroborates the result reported for $LaMn_{0.5}Fe_{0.5}O_3$ phase,[8] which has been explained by the local magnetic ordering below $T_C$ instead of bulk FM ordering. Our investigations also support this view point. The origin of FM seems to be due to the positive interactions between $Fe^{3+}$ and $Mn^{3+}$ ions (may be also between $Mn^{3+}$ and $Mn^{4+}$). The unsaturated behaviour of M(H) curves, even at high fields, suggests the presence of a glassy-FM state at low temperature.[7,12]

In order to further establish the glassy behavior, frequency dependent magnetic measurements were performed at low temperature. Figure 1(b) shows the temperature dependent in-phase, $\chi'(T)$, component of the ac-susceptibility measured at different frequencies ($h_{AC}$=10Oe). The $\chi'(T)$ data exhibited similar features to the ZFC magnetization data. The system shows a weak anomaly corresponding to FM $T_C$, which is frequency-independent and a frequency-dependent broad maximum corresponding to ZFC cusp (around 28K). The latter shifts towards higher temperature with increasing frequency (see inset of Fig. 1b), which is a characteristic feature of spin-glasses.[12-13] From the above results, it clearly appears that the FM and AFM components are competing at low temperature, due to the presence of $Mn^{3+}$ and $Fe^{3+}$ ions akin to $LaMn_{0.5}Fe_{0.5}O_3$ system.[8] The $Fe^{3+}$-O-$Mn^{3+}$ super exchange interactions are indeed responsible for the FM component, characterized by a finite value of coercive field with hysteresis loops. Nevertheless, the obtained $\mu_{eff}$ value (~5.49 $\mu_B$/f.u.) is smaller than the theoretical spin-only value for $Fe^{3+}$ and $Mn^{3+}$ ions in high spin states and the lower value of paramagnetic Curie temperature ($\theta_p \sim 51K$) than $T_C$, clearly indicate that there are strong AFM interactions present in the sample, due to $Fe^{3+}$-O-$Fe^{3+}$ and $Mn^{3+}$-O-$Mn^{3+}$ interactions. Consequently, the system might be electronically phase separated into FM and AFM domains, leading to a competition between these two interactions giving rise to a glassy-FM state.[8,12]

In order to investigate the magneto-dielectric and/or FE property, dielectric and pyroelectric measurements were carried out as a function of temperature and magnetic field using a Physical Properties Measurements System (PPMS, Quantum Design) coupled with an impedance analyzer (Agilent technologies-4284A) in the temperature range of 10-300K. The electrodes were prepared in capacitor geometry with either side of the polished pellets painted with silver paste (Dupont). Fig. 2. shows the dielectric behavior of the pellets measured in the frequency range of 10 kHz – 1 MHz. The $\varepsilon'$ (real part of dielectric constant) is observed to vary from 20 – 600 and the $\varepsilon''$ (imaginary part of dielectric constant) varied from 0.01 - 6000 in the measured temperature range. The rise in the dielectric constant to a colossal magnitude around 120K associated with a local peak in the imaginary dielectric constant (indicated with a downward arrow around 160K in Fig. 2) could be due to extrinsic effects such as grain boundaries (Maxwell-Wagner type), twin boundaries (if any present in the system) and/or other conductive effects.[13-14] Above 230K, the electrode material interface and space charge effects due to the semiconducting behavior of the sample, collectively dominate the capacitance measurements and display a monotonous rise in the magnitudes of both real and imaginary permittivity.[15] The electrical resistance measurement in the temperature range of 100-400K, also depicts semiconducting behavior near the room temperature and at low temperature resistance increases exponentially akin to insulator. The impedance spectroscopy studies or samples with varying thickness and electrodes[17] (which is beyond the scope of this article) could assist isolating the individual components dominating at different temperature and frequency region.[13] A weak positive magneto-dielectric effect of around 0.15-0.25% observed in the temperature range of 10-80K, which, is expected to be an intrinsic effect of the sample.[3] Whereas, a large magneto-dielectric effect of around 0.3-10% is observed in the range of 100-300K with a maximum around 220-250K, which could arise due to other extrinsic effects, such as grain boundary conductance.[13] The inset of Fig. 2 shows a magnified view of $\varepsilon'$ in a narrow temperature range of 50-72K. A weak anomaly in the $\varepsilon'$ was observed at ~ 62K and at ~ 65K in the $\varepsilon''$ (not shown here) on repeated measurements. The anomaly is predominant and highly reproducible in the presence of magnetic field as shown in the inset Fig. 2, and could originate from the onset of polar behavior. A recent theoretical study on magnetic perovskite, reveals that the plausible polar ground states can couple with an octahedral lattice, consisting of different magnetic cations.[6] The presence of $6s^2$ lone pair electrons in the $Bi^{3+}$ cation is expected to introduce an additional structural distortion into the lattice. The latter can explain the anomaly observed in the dielectric studies. Low temperature polarization measurements were performed to confirm this..

The polarization measurements were carried out from the pyroelectric current measurements.[16] The sample was heated to 300K and then cooled down to 120K with zero electric field (due to the semiconducting nature above 120K) and further cooled down to 10K in the presence of electric field. On the removal of electric field, the capacitor was discharged in the zero electric field, in order to remove the stray charges in the circuitry and the accumulated charges in the sample.[17] Later, the pyroelectric current was measured in the temperature range of 10 - 80K and the same procedure is followed for different field values.[16] Fig. 3a shows the temperature dependent electric polarization in an applied electric field of ±3.2kV/cm, which depicts a clear polar behavior with a remnant polarization of 0.3μC/cm$^2$ at 10K. A polar-to-non polar kind of transition is evidenced around 67K. The measured polarization values increased with the applied electric field and saturated (0.3μC/cm$^2$) at applied electric fields above 3.2kV/cm. Due to the experimental limitations in measuring the ferroelectric hysteresis, the reversibility of the polarization with applied electric field was confirmed in the similar way of the polarization measurements. The reversal of polarization with applied field is further investigated (Fig. 3b). The sample was cooled down from 300K to 50K in an applied electric field of +3.2 kV/cm. The capacitor was then discharged at 50K, for more than 3x10$^3$ s. Later the polarization was measured from 50 to 40K (step 1 in Fig. 3b). Then, a field value of -3.2kV/cm is applied at 40K for observing a reversal of polarization (-0.3 μC/cm$^2$). The capacitor was discharged again at 40K for 3x10$^3$ seconds followed by the polarization measurement from 40-30K as explained earlier (step 2). The similar experiment was performed at 30K with +3.2kV/cm again to observe the reversal of polarization (to 0.3 μC/cm$^2$) : a clear reversal of polarization is observed successively down to 20K (step 3). Nevertheless, on application of -3.2kV/cm at 20K a complete reversal of polarization to negative value was not observed (step 4), which could be due to the thermally pinned polar clusters present associated with the freezing of magnetic domains at low temperature. This was further confirmed by the complete reversal of polarization reversal observed on applying higher field values (-5.5 kV/cm) at 20K, as shown in Fig.3b (step 5). The observed polarization behavior and its reversal nature, was noticed both in the heating and cooling cycles of the sample. Above 80K, the measurement was dominated by the thermally activated current and leaves the true polarization immeasurable. The dielectric anomaly associated with a drop in polarization reveals the polar behavior of the samples below 67K. The observed polarization is also reversible with applied field and effectively satisfies the ferroelectric criteria.[17] Finally, the coexistence of both ferromagnetic and ferroelectric behavior proves the multiferroic nature exhibited by LBMFO.

In summary, the simultaneous existence of a ferromagnetic state and of a polar state for $La_{0.5}Bi_{0.5}Mn_{0.5}Fe_{0.5}O_{3.09}$ and especially the reversal of its polarization under application of an electric field, demonstrate that this oxide is a multiferroic. This study suggests that many perovskites $ABO_3$ which exhibit separately FM interaction between the B-cations of their octahedral $BO_3$ lattice, and local structural distortions induced by lone pair cations on their A-sites, present a great potential to realize multiferroic features.

We gratefully acknowledge the CNRS, the French Ministry of Education and Research and the European project MaCoMuFi (NMP3-CT-2006-033221), for financial support.

**Figure captions:**

**Figure 1.** (color online) (a) (M-H) loops recorded at several temperatures. (b) Temperature dependent in-phase component of magnetic ac-susceptibility, χ'(T), for different frequencies ($h_{AC}$ = 10 Oe). Inset shows the enlarged version around ZFC cusp (near 28K) for LBMFO.

**Figure 2.** (color online) Temperature dependent of the real ($\varepsilon'$) and imaginary ($\varepsilon''$) parts of permittivity in different magnetic fields. Inset shows the magnified view near the weak dielectric anomaly (around 62K).

**Figure 3.** (color online) (a) Temperature-dependence of electric polarization in an applied field of ±3.2kV/cm. (b) Electric field dependent polarization reversal at low temperature on different applied field directions. The arrows and number of steps indicate the way the measurement is made, see text for details (solid lines are guide to the eyes).

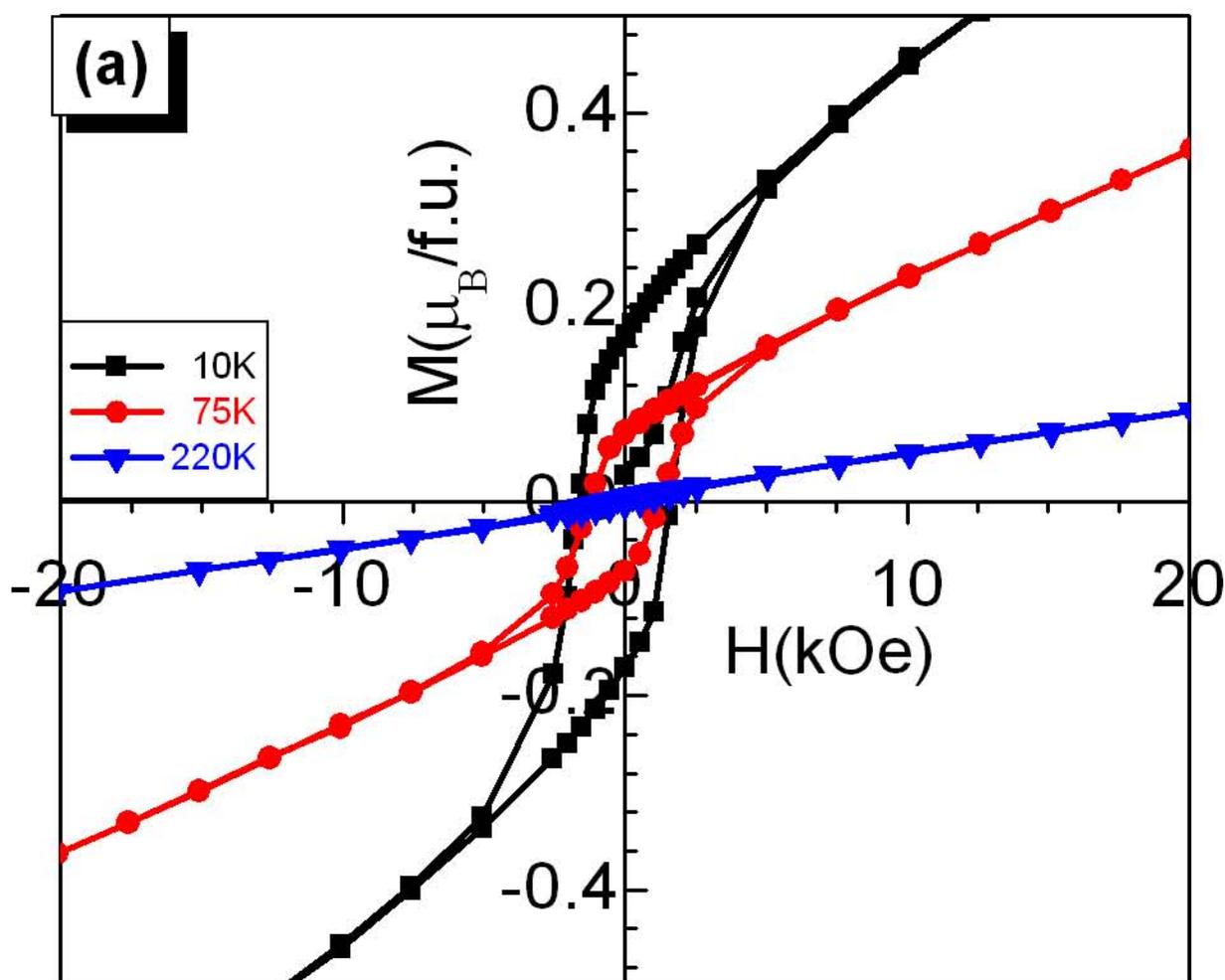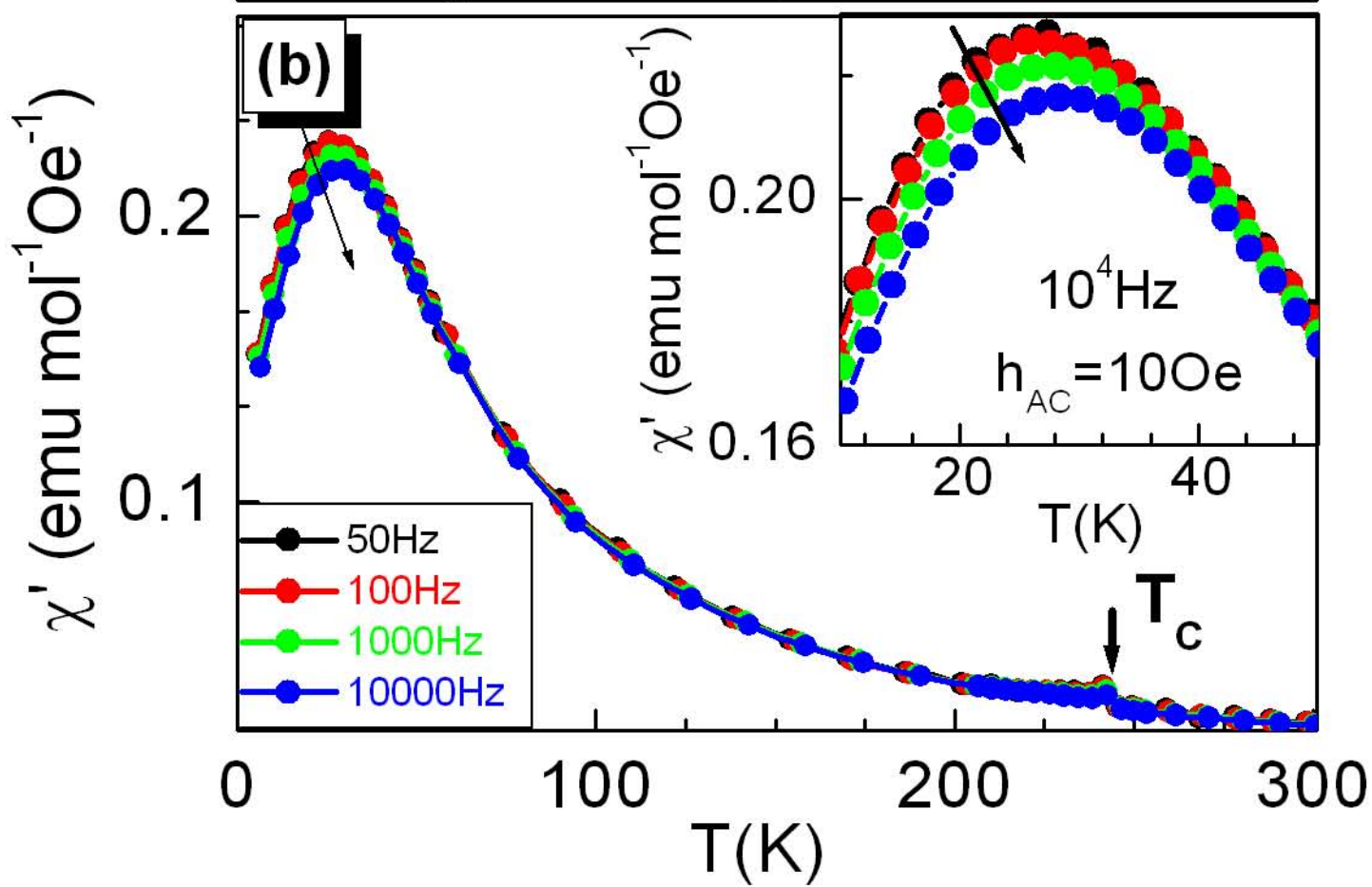

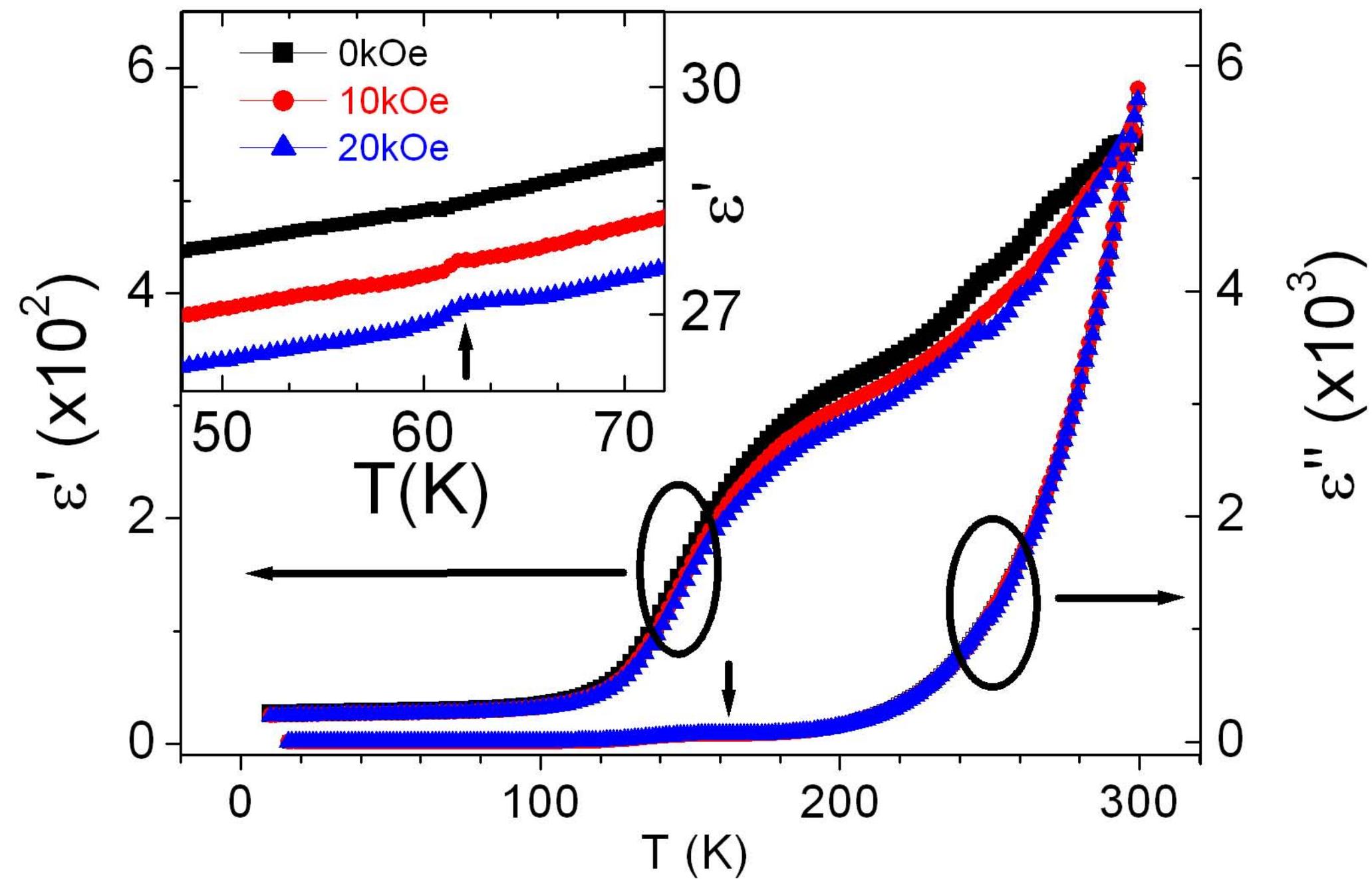

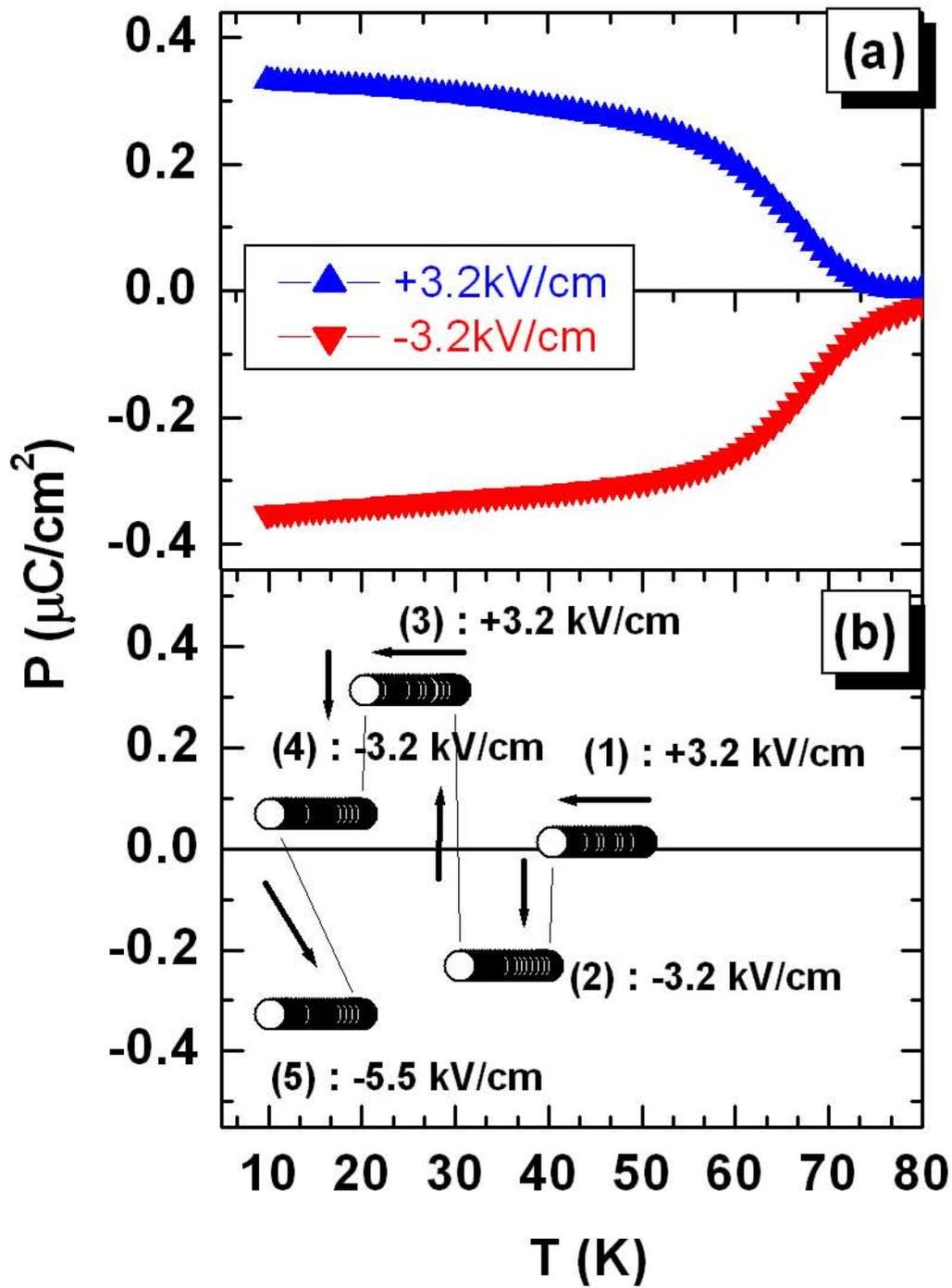